\documentclass{aa}  
\usepackage{graphicx}
\usepackage{txfonts}
\usepackage{xcolor}
\usepackage[breaklinks, colorlinks, citecolor=blue]{hyperref}
\newcommand{\Planck}{{\it Planck}}

\newcommand\horsp{\rule[-3mm]{0mm}{8mm}}
\defcitealias{Tanimura2021}{T21}

\begin{document} 

   \title{Retrieving cosmological information from
   small-scale \\CMB foregrounds}

   \subtitle{I. The thermal Sunyaev Zel'dovich effect}

   \author{Marian Douspis\inst{1}
          \and
          Laura Salvati\inst{1,2}
          \and 
          Adélie Gorce\inst{3}
          \and 
          Nabila Aghanim\inst{1}
          }

   \institute{
    Université Paris-Saclay, CNRS,  
    Institut d'Astrophysique Spatiale, 
    91405, Orsay, France
    \and 
    INAF – Osservatorio Astronomico di Trieste, Via G. B. Tiepolo 11, 34143 Trieste, Italy
    \and
    Department of Physics and McGill Space Institute, McGill University, Montreal, QC, Canada H3A 2T8 \\
    \email{marian.douspis@universite-paris-saclay.fr}
             }
\date{Received *********; accepted *******}

 
 \abstract{ We propose a new analysis of small-scale cosmic microwave background (CMB) data by introducing the cosmological dependency of the foreground signals, focussing first on the thermal Sunyaev-Zel'dovich (tSZ) power spectrum, derived from the halo model. We analyse the latest observations by the South Pole Telescope (SPT) of the high-$\ell$ power (cross) spectra at 95, 150, and  220 GHz, as the sum of CMB and tSZ signals, both depending on cosmological parameters and remaining contaminants. In order to perform faster analyses, we propose a new tSZ modelling based on machine learning algorithms (namely Random Forest). We show that the additional information contained in the tSZ power spectrum tightens constraints on cosmological and tSZ scaling relation parameters. We combined for the first time the {\it Planck} tSZ data with SPT high-$\ell$ to derive new constraints. Finally, we show how the amplitude of the remaining kinetic SZ power spectrum varies depending on the assumptions made on both tSZ and cosmological parameters. These results show the importance of a thorough modelling of foregrounds in the cosmological analysis of small-scale CMB data. Reliable constraints on cosmological parameters can only be achieved once other significant foregrounds, such as the kinetic SZ and the cosmic infrared background (CIB), are also properly accounted for.}

  

\maketitle
%

\section{Introduction}

Observations in millimetre wavelengths for  cosmic microwave background (CMB) analysis contain, at small (arcmin) scales, a sum of signals originating from sources ranging from the last scattering surface to our local environment. At these scales, the primordial CMB signal fades away compared to secondary anisotropies and contaminants created along the path of photons. 
Among these, the thermal and kinetic Sunyaev-Zel'dovich effects \citep[tSZ and kSZ, respectively ;][]{zeldovich_sunyaev_1969,sunyaev_zeldovich_1980}, which stem from the interaction between CMB photons and the large-scale structure of the recent Universe, contain information of cosmological interest.

The inverse Compton effect of CMB photons on energetic electrons from hot gas in galaxy clusters and groups (and thus filaments) is responsible for the tSZ anisotropies and additional power at small scales in the angular power spectrum of CMB fluctuations. Such an effect is frequency-dependent, a property which allows one to separate tSZ signal from other components in CMB data and reconstruct both a tSZ map \citep{Planck13, Planck15_tsz_map, PACT, ACTmap, SPTmap, Tanimura2021} and power spectrum \citep{Planck13, Planck15_tsz_map, Tanimura2021}.  
The power spectrum amplitude and shape of this secondary anisotropy are highly dependent on the number of halos and their distribution in mass and redshift, and thus on the cosmological model considered. Typically, the amplitude of the tSZ spectrum varies with the normalisation of the matter power spectrum as $\sim \sigma_8^{8.37}$ and the total matter content of our Universe as $\sim \Omega_m^{3.29}$ \citep[see, e.g.][]{HurierLacasa,Salvati2018,Bolliet18}. 
As the cluster mass is never observed directly, the tSZ amplitude also depends on the modelling and calibration of the scaling relation (SR) between the mass and the observable, that is the gas pressure. 

On the other hand, the kSZ effect is produced by induced Doppler shift of the scattered CMB photons off free electrons with a bulk velocity with respect to the CMB rest-frame. It has the same frequency dependency as CMB photons. Due to this interaction, anisotropies are created at two epochs: first, during the reionisation era, as the effect depends on the number and distribution of free electrons, and at later times in large-scale structures -- typically, clusters \citep{Aghanim96, ACTkSZ}. These anisotropies translate into an additional power spectrum, peaking at small scales, for which the shape and amplitude depends on the reionisation history \citep{ZahnZaldarriaga_2005, Planck2015_reion,Gorce2020}.  At the same millimetre wavelengths, galaxies may contribute to the observed signal by emission in the radio  (synchrotron of active galactic nuclei) or infrared (dusty galaxies) domains, and even dominate at smaller scales. \\

Most current analyses of CMB observations at small scales are done following the same approach  \citep[see, e.g.,][]{George:2014oba, 2017A&A...597A.126C, Reichardt2020, 2021JCAP...01..031H}. A theoretical CMB power spectrum is added to templates of each non-CMB signal to reproduce the observed power spectrum. As mentioned above, apart from the tSZ and kSZ effects, the remaining non-CMB components are thermal dust emission from dusty star-forming galaxies (DSFG) -- both the Poisson and spatially clustered component, the radio galaxy emission, the Galactic cirrus signal and the cosmic infrared background (CIB). The cosmological parameters enter only the CMB power spectrum, whilst the amplitudes of all templates are set free and marginalised over:
\begin{equation}\label{eq:clsobs}
C_{\ell}^{\text{obs}} \equiv C_{\ell}^{\text{CMB}} (\Theta) + \sum_i D_{3000,i} \times T_i \, ,
\end{equation}
where $\Theta$ is the set of free cosmological parameters and $D_{3000,i}$ is the free amplitude of a template $T_i$ for foreground component $i$, normalised to one at $\ell=3000$. All but CMB signals are considered nuisance  quantities and the cosmological information is retrieved only from the CMB part\footnote{A similar approach allows also to set the amplitude as depending on typical approximated scaling relations between the amplitude of the signal at $\ell=3000$ and some cosmological parameters, e.g. $A_{\text{tSZ}} \propto \sigma_8^{9}$, but loses any  information coming from the shape.}. Little to no astrophysical information can be derived from the secondary anisotropies, unless it is directly contained in the amplitude of their spectra. Such analyses may also include inconsistencies in the assumptions made in the different templates and in the CMB computation: a typical example is the cosmological parameters assumed in the simulation used to derive the tSZ template that may be different from the one used for the kSZ template, and from the resulting, best-fit cosmological parameters -- see, for example, discussions in  \citet{George:2014oba} and \citet{Reichardt2020}.

Our goal, here, is to use the full cosmological information contained in both primordial and secondary anisotropy power spectra, in order to get tighter constraints on the main cosmological parameters and at the same time on the physical parameters related to secondary anisotropies. 
Our approach, as in \citet{2006A&A...456..819D}, is to substitute templates of the non-CMB signal with 
models for the power spectra that encompass the full cosmological information .

As a first step, we focus on the tSZ effect, proposing also a new way to compute efficiently its angular power spectrum at high $\ell$, including it in the traditional Monte Carlo Markov Chains approach. Doing so, we are able to put constraints on cosmological parameters, cluster models and residual kSZ amplitude:
\begin{equation}
C_{\ell}^{\text{obs}} \equiv C_{\ell}^{\text{CMB}} (\Theta) + C_{\ell}^{\text{tSZ}}(\Theta, \Sigma)+ \sum_j D_{3000,j} \times T_j \, , 
\end{equation}
where $\Sigma$ is the set of scaling relation parameters and $j$ runs over all the remaining non-CMB components. 

We apply our approach to the South Pole Telescope (SPT) data presented in \citet{Reichardt2020}, which spans a range of scales and frequencies where the SZ signal dominates the primordial CMB, and to the \Planck-tSZ data of \cite{Tanimura2021}. We investigate the impact of this new approach on cosmological, scaling relation parameters and kSZ amplitude constraints. 
A companion paper will present how, in addition,  the kSZ template could be substituted by a more physical modelling, presented in \cite{Gorce2020}, to derive information on the Epoch of Reionisation \citep*{PAPER3}. 

The next section details the ingredients required to compute the tSZ power spectrum from the halo model and introduces the scaling relation (SR) parameters. A sub-section presents a new technique introduced in this paper to speed up the computation of the tSZ power spectrum, based on machine learning. Section~\ref{sec:likelihood} introduces the SPT and \Planck-tSZ data used in our analysis and the details of the MCMC settings. Section~\ref{sec:results} shows the changes in constraints obtained with our approach on cosmological and SR parameters. Finally, we discuss how the significance of the resulting kSZ power amplitude may change depending on the assumptions on the tSZ contribution.
The assumed cosmological and scaling parameters and their respective priors are summarised in Appendix \ref{app:params}.

\section{The tSZ angular power spectrum}
 \label{sec:tSZ}

In this section, we present the adopted theoretical model for the tSZ power spectrum and describe the machine learning method used to evaluate the tSZ signal in the MCMC sampling.

\subsection{Theoretical evaluation}\label{sec:CLtSZ_th}

For the theoretical evaluation of the tSZ power spectrum, we follow the approach discussed in \citet{Planck15_tsz_map} and \citet{Salvati2018}. For the complete discussion, we refer the reader to these works and report here only the main steps of the derivation.

We adopt the halo model \citep[see, e.g., ][]{Cooray2000}, and write the total tSZ power spectrum as the sum of a one-halo and a two-halo term
\begin{equation}\label{eq:Cl_tot}
    C_{\ell}^{\rm tSZ} = C_{\ell}^{1 \rm halo} + C_{\ell}^{2 \rm halo} \, .
\end{equation}
Under the flat-sky approximation, the one-halo term is defined as 
\begin{eqnarray}\label{eq:Cl_1halo}
    C_{\ell}^{1 \rm halo} &=& \int _0^{z_{\rm max}} dz \, \dfrac{dV}{dz d\Omega} \notag \\
    & \times & \int_{M_{\rm min}}^{M_{\rm max}} dM \, \dfrac{dn(M_{500,z})}{dM_{500}} |\tilde{y}_{\ell} (M_{500},z)|^2. 
\end{eqnarray}
In the above equation, $dn(M_{500,z})/dM_{500}$ is the halo mass function, and $dV/(dz \, d\Omega)$ is the comoving volume element (per unit of redshift and solid angle). We assume $M_{500}/[M_{\odot} h^{-1}] \in [10^{13},~5 \times 10^{16}]$, $z\in[0,3]$.
For the halo mass function, we adopt the formulation described in \citet{TinkerKravtsov_2008}.
The term $ \tilde{y}_{\ell} (M_{500},z)$ represents the Fourier transform, on a sphere, of the Compton parameter $y$. Following the Limber approximation, it is defined as
\begin{eqnarray}\label{eq:y_ell}
    \tilde{y}_{\ell} (M_{500},z) &=& \dfrac{4 \pi r_s}{\ell_s^2} \left( \dfrac{\sigma_T}{m_e c^2} \right) \notag \\
    & \times & \int_0^{\infty} dx \, x^2 P_e(M_{500},z,x) \dfrac{\sin{(\ell_x/\ell_s)}}{\ell_x/\ell_s} \, .
\end{eqnarray}
For the pressure profile $P_e(M_{500},z,x)$, we adopt the universal formulation provided in \citet{ArnaudPratt_2010}, with $r_s$ being the scale radius and $x=r/r_s$. The multipole $\ell_s$ is defined from the angular diameter distance $D_A(z)$, such that $\ell_s = D_A(z)/r_s$. 

The two-halo term can be defined as
\begin{eqnarray}\label{eq:CL_2halo}
    C_{\ell}^{2 \rm halo} &=& \int _0^{z_{\rm max}} dz \, \dfrac{dV}{dz d\Omega} \notag \\
    & \times & \left[ \int_{M_{\rm min}}^{M_{\rm max}} dM \, \dfrac{dn(M_{500,z})}{dM_{500}} \, \tilde{y}_{\ell} (M_{500},z) \, B(M_{500},z) \right] ^2 \notag \\
    & \times & P(k,z) \, .
\end{eqnarray}
In the above equation, $P(k,z)$ is the matter power spectrum and $B(M_{500},z)$ is the time-dependent linear bias factor
which relates the correlation function of the tSZ signal to the underlying matter power spectrum. Following \citet{KomatsuKitayama_1999}, we define the bias as 
\begin{equation}\label{eq:bias_2halo}
    B(M_{500},z) = 1 + \dfrac{\nu^2(M_{500},z)}{\delta_c(z)} \, ,
\end{equation}
where $\nu(M_{500},z) = \delta_c(M_{500})/D(z)\,\sigma(M_{500})$. Here, $D(z)$ is the linear growth factor, and $\delta_c(z)$ is the over-density threshold for the spherical collapse. The term $\sigma(M_{500})$ is the standard deviation of density perturbations, at present time, in a sphere of radius $R = (3M_{500}/4\pi \rho_{m,0})$, defined as
\begin{equation}\label{eq:sigma_Mz}
    \sigma^2(M) = \frac{1}{2 \pi^2} \int dk \, k^2 \, P(k,z) \, |W(kR)|^2 \, ,
\end{equation}
where $W(kR)$ is the window function of a spherical top-hat of radius $R$.
In the remaining of this work, we may refer to the $C_{\ell}^{\text{tSZ}}$ as $C_{\ell}$ when there is no ambiguity, but always put attributes when we refer to other components (CMB, kSZ, CIB, ...).

We also take into account the non-Gaussian contribution to the cosmic variance, defined by the trispectrum $T_{\ell \ell'}$. Following \citet{Komatsu2002} and \citet{Horowitz2017}, the dominant term of the trispectrum in the halo model is
\begin{eqnarray}\label{eq:Tll}
    T^{\text{tSZ}}_{\ell \ell'} &\simeq & \int_{0}^{z_{\rm max}} dz \, \dfrac{dV}{dz \, d\Omega} \notag \\
    & \times & \int_{M_{\rm min}}^{M_{\rm max}} dM \, \dfrac{dn(M_{500,z})}{dM_{500}} \, |\tilde{y}_{\ell} (M_{500},z)|^2 \,  |\tilde{y}_{\ell'} (M_{500},z)|^2 .
\end{eqnarray}

\subsection{Scaling relations}\label{sec:SR}

As mentioned above, the halo mass cannot be measured directly. We need, therefore, to define a scaling relation between the chosen observable and the mass. For the tSZ effect, we can define this relation starting from the integrated Compton-$y$ parameter, $Y_{500}$. Following \cite{Planck2015_tsz_map}, it reads:
\begin{equation}\label{eq:sr_ym}
    E^{-\beta}(z) \left[ \dfrac{D_A(z)^2 Y_{500}}{10^{-4} \text{Mpc}^2} \right] = Y_\star \left[ \dfrac{h}{0.7} \right]^{-2+\alpha} \left[ \dfrac{(1-b) M_{500}}{6 \cdot 10^{14} M_{\odot}} \right]^{\alpha} \, ,
\end{equation}
with a dispersion $\sigma_{\ln Y}=0.17$. 
In the above equation, $D_A(z)$ is the angular diameter distance, and $E(z) \equiv H(z)/H_0$. The parameters $\beta, \, Y_*, \, \alpha, \, (1-b)$ are obtained by calibrating this scaling relation with external multi-wavelength data. In particular, we can assume that clusters follow a self-similar evolution, which implies $\beta=2/3$. The $Y_*$ and $\alpha$ parameters are constrained from X-ray data. 
The parameter $b$ represents the mass bias and it is related to the ratio between the tSZ mass, evaluated from the pressure profile assuming hydrostatic equilibrium, and the real cluster mass $M_{500}$, such that $(1-b) \equiv M_{\text{SZ}}/M_{500}$. This mass bias is calibrated with weak lensing observations. We refer the interested reader to \citet{Planck2015_tsz_map} for additional details of the definition and general calibration process of the scaling relation.

\subsection{Machine learning approach}\label{sec:RF}

The derivation of the tSZ power spectrum as described in Sec.~\ref{sec:CLtSZ_th} implies the computation and integration of the Fourier transform of the Compton-$y$ parameter over a large range of masses, redshifts and $k$-modes. As the multipole range increases towards higher $\ell$ with high resolution instruments like SPT, the time to compute the $C_{\ell}^{\text{tSZ}}$'s increases and slows down the sampling. Therefore, we turn to machine learning techniques to approximate the full $C_\ell$'s computed above (hereafter `true $C_\ell$'s') to speed up the MCMC runs (see also \citet{2021arXiv210603846S} for the development of an emulator for CMB and large-scale structure analyses). 

Among the numerous existing machine learning techniques, we choose a Random Forest (RF) algorithm. Random Forest is a supervised machine-learning algorithm which bootstraps the training set in order to construct independent decision trees. The final prediction is the mean of the outputs of the different trees.
The training set is divided into a learning set, a validation set and a testing set, constructed to evaluate potential bias and errors in the prediction.
Random Forest has generated great interest in recent astrophysical and cosmological analyses \citep[e.g.][]{Bonjean2019, RFex1,RFex2}, because it achieves high accuracy and efficiency with large data sets, whilst being quick to implement.
Additionally, at the training stage, RF is able to learn in a fast way highly non-linear relations between the inputs and the outputs it is given.
For these reasons, we implement this particular machine-learning technique in our analysis.\\

  \begin{figure}
   \centering
   \includegraphics[width=\columnwidth]{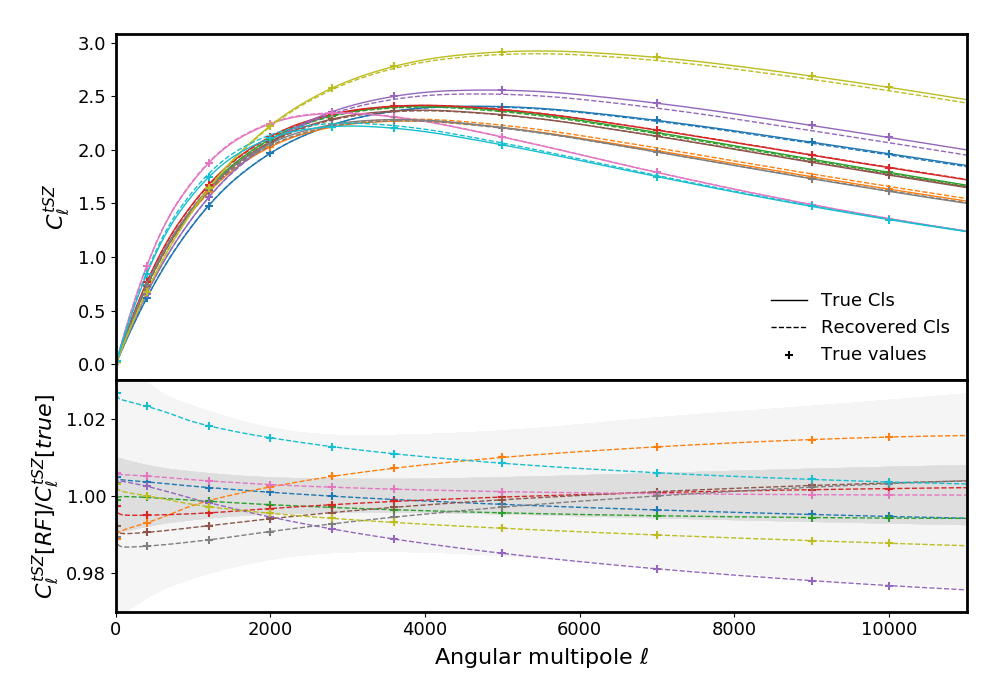}
      \caption{Comparison of the `true' tSZ $C_\ell$'s values at given $\ell$ (crosses, and solid lines) with the RF-inferred $C_\ell$'s after interpolation (dashed lines) for a sample of models (units are arbitrary before renormalisation in $y$-units at 143~GHz). The lower part of the plot shows the ratio of RF $C_\ell$'s over `true' $C_\ell$'s. The grey areas are showing the 68\% and 95\% dispersion among the test set.}
    \label{fig:Cls}
   \end{figure}
   
We build a training set with a random sampling of an eight-dimensional space of cosmological and scaling relation parameters as inputs: five cosmological parameters ($\Omega_b$, $\Omega_m$, $\sigma_8$, $n_s$ and $h$) and three scaling relation parameters ($1-b$, $Y_\star$, $\alpha$), on a wide range including their fiducial values (see last column of Table~\ref{tab:paramstraining} for precise ranges).
The outputs consist of the corresponding $C_\ell$'s computed at 10 different $\ell$-bins between 10 and 10~000. 
Given the smoothness of the power spectrum, we find this number of bins to be sufficient to recover by interpolation the $\ell$-by-$\ell$ power spectrum required by the likelihood for the cosmological analysis.

We use a sample of 15~000 models divided into fifths such that 60\%, 20\% and 20\% of the models are used for the learning, validation, and testing phase, respectively. In order to facilitate the learning, we pre-process the  $C_\ell$'s by rescaling them accordingly to their order zero amplitude variation at $\ell=3000$ with the cosmological and scaling parameters. For example, the learning is done on $C_\ell /\Omega_m^{3.29}$ instead of $C_\ell$. 
Doing so, the RF only needs to learn the shape and variation around $ \Omega_m^{3.29}$ instead of the full dependency.  We generalise this procedure to all considered parameters, such that the RF learning phase is done on $C_\ell^{\text{tSZ}}/C_{\rm dep}^{\text{tSZ}}$ with :
\begin{equation}
\begin{aligned}
    C_{\rm dep}^{\text{tSZ}} = \;\Omega_{m}^{3.29} \;  \sigma_8^{8.37} \; n_s^{0.60} \; &h^{0.09} \; \Omega_b^{-0.0005} \\
    & \times \; (1-b)^{2.96} \; \alpha_{SZ}^{-2.19} \; Y_\star^{1.83} .
    \end{aligned} 
\end{equation} 
The exponent of each parameter is obtained by minimising the spread of the $C_{\ell=3000}^{\text{tSZ}}/C_{\rm dep}^{\text{tSZ}}$ distribution for all the parameters in the training set.
We find that pre-processing the data in this way improves the score of the prediction by 5\%.

To further improve the accuracy of the prediction, a possibility is to optimise the choice of RF hyper parameters.
In this perspective, we test different numbers of trees and depths and choose, respectively, 40 and 20, as they maximise the final score of the prediction to 96.2\%\footnote{We call score the coefficient of determination $R^2$ of the prediction. The best possible score is $100\%$.}, whilst limiting memory and time imprints.

An example of the quality of the RF-inferred $C_\ell$'s versus `true' ones is shown for a sample of ten random models, picked within the testing set, in Fig.~\ref{fig:Cls}. This technique reproduces well the true $C_\ell$'s with an overall gain in computation time of about two orders of magnitude. To further improve the computation time, one could use another emulator to predict the CMB power spectrum instead of using a Boltzman integrator in the analysis \citep{2021arXiv210603846S}. We find that the true $C_\ell$'s and the RF $C_\ell$'s overall agree at better than the 2\% level. As seen on Fig.~\ref{fig:Cls}, this corresponds to $0.08~\mu \mathrm{K}^2$ at $\ell=3000$, which is below the error of current  measurements: $ \sim 1.08~\mu \mathrm{K}^2$ at $\ell=3000$ at $2\sigma$ \citep{Reichardt2020}. However, a better reconstruction might be required for future observations. 
 
\section{Data and method}
\label{sec:likelihood}

In this section, we describe the different datasets used in our analysis and the approach we adopt for the MCMC sampling of the parameter space.

\subsection{SPT 2020}

We consider the observed  signal on small scales (high multipoles $\ell$) and, therefore, make use of the SPT temperature-only data and likelihood introduced in \citet{Reichardt2020} and made publicly available by the authors\footnote{Available at \url{https://pole.uchicago.edu/public/data/reichardt20/}.}. We refer the interested reader to this work for details on the dataset, but highlight the fact that the data considered spans the range $2000\leq \ell \leq 11000$ at frequencies 95, 150 and 220~GHz. The likelihood makes use of auto- and cross-spectra and marginalises over calibration and beam parameters. In the baseline analysis of \citet{Reichardt2020}, the total signal is modelled by CMB, tSZ, kSZ, galactic cirrus contamination, radio and infrared galaxies spectra and tSZ-CIB cross-spectra. In this work, we modified only the tSZ spectra and consequently the tSZ-CIB cross spectra contributions, computed in the SPT likelihood as the sampled correlation coefficient times a function of the tSZ $C_{\ell}$'s \citep[as defined in][]{Zahn12}. Implementing new CIB and kSZ modelling will be the focus of upcoming works.

\subsection{\Planck\ tSZ}\label{sec:Planck_tSZ}

To complement the high-$\ell$ SPT data, we consider the \Planck21-tSZ power spectrum estimated by \citet[][hereafter T21]{Tanimura2021}, an update of  the \Planck\ collaboration  analyses \citep{Planck13, Planck2015_tsz_map} but  using the
\Planck\ frequency maps from the public data release PR4 \citep{2020A&A...643A..42P}. 
The scales covered by \Planck21-tSZ are $60 < \ell < 1400$ and do not overlap with those covered by SPT. Hence, there is no need to consider correlations between the two experiments when combining the data sets. When using \Planck21-tSZ data, we consider the map power spectrum and marginalise over residual foregrounds amplitudes (CIB, IR and radio point sources, and residual high-$\ell$ noise), as done and described in \citetalias{Tanimura2021}, such that
\begin{equation}
C_{\ell}^{\rm obs} \equiv  C_{\ell}^{\rm tSZ}(\Theta, \Sigma)+ A_{\rm CIB}C^{\rm CIB}_{\ell}+A_{\rm IR}C^{\rm IR}_{\ell}+A_{\rm Rad}C^{\rm Rad}_{\ell}+A_{\rm res.}C^{\rm res.}_{\ell}.
\end{equation}

We thus build our \Planck21-tSZ likelihood, assuming Gaussian priors on the foreground dimensionless amplitudes centred on one and with a standard deviation of 0.5 \citepalias{Tanimura2021}. In addition to the statistical error bars, we also consider the error term coming from the trispectrum contribution and  defined as \citep[see, e.g.,][]{Komatsu2002,Horowitz2017}
\begin{equation}\label{eq:sigma_Tll}
    \sigma_{T_{\ell \ell'}} = \dfrac{\ell (\ell +1)}{2 \pi} \left( \dfrac{T_{\ell \ell'}}{4 \pi f_{\rm sky}} \right),
\end{equation}
where we use $f_\mathrm{sky} = 0.42$, to be consistent with T21 analysis.
To ease the computation while using the RF $C_\ell$'s, we approximate the $\ell$-by-$\ell$ trispectrum term $\sigma_{T_{\ell \ell'}}$ with a constant value over the full multipole range. We consider the mean value of Eq.~\eqref{eq:sigma_Tll}, evaluated at the fiducial cosmology. We keep this value independently of the cosmological model considered, as we made sure that this approximation only introduces a negligible uncertainty with respect to the total error budget. Additionally, in Sec.~\ref{sec:results}, we show that this approximation has no impact on the inference of cosmological and scaling relation parameters.

  \begin{figure}[!ht]
   \centering
   \includegraphics[width=8cm]{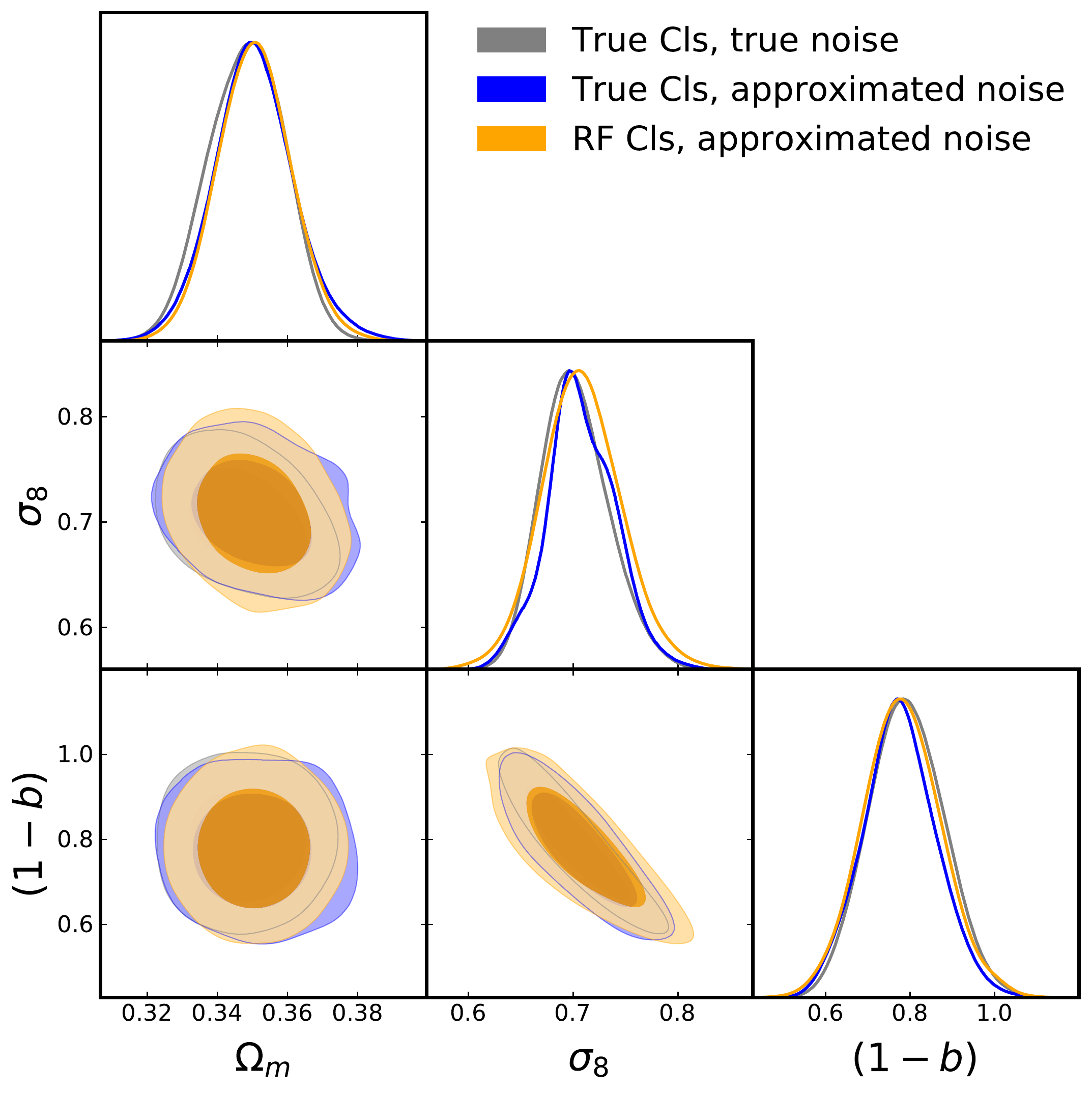}
      \caption{Comparison of MCMC runs with `true' $C_\ell$'s and RF $C_\ell$'s for cosmological and scaling relation parameters. In grey, both $C_\ell$'s and the non-Gaussian contribution ($\sigma_{T_{\ell \ell'}}$) are computed as detailed in Sec.~\ref{sec:CLtSZ_th}. In blue,  $\sigma_{T_{\ell \ell '}}$ is approximated by a constant value over $\ell$ and cosmology. In orange, the $C_\ell$'s are computed with the RF and the sample variance noise is approximated by a constant.}
    \label{fig:verif}
   \end{figure}

\subsection{MCMC}

We make use of the publicly available Monte Carlo Markov Chain \texttt{CosmoMC} code \citep{cosmomc,cosmosmc2}, that we modify to include both the SPT high-$\ell$ and the \Planck21-tSZ likelihoods, as well as the RF prediction of the tSZ power spectrum. 
We sample simultaneously cosmological and scaling relation parameters. For the cosmological model, we consider five standard parameters, that is the baryon and cold dark matter densities, $\Omega_b h^2$ and $\Omega_c h^2$; the ratio of the sound horizon to the angular diameter distance at decoupling $\theta$; the scalar spectral index, $n_s$; the overall normalisation of the spectrum, $A_s$;  the reionisation optical depth $\tau$ is assumed fixed. For the scaling relations, we follow the discussion in Sec.~\ref{sec:SR} and, therefore, sample three parameters, that is the mass bias $(1-b)$, the mass slope $\alpha$ and the normalisation $Y_*$. As stated in Sec.~\ref{sec:SR}, we assume the self-similar evolution for halos, and fix $\beta=2/3$.

We focus our analysis on the following cosmological parameters: $\Omega_m$, $\sigma_8$, $H_0$ , and the mass bias parameter $(1-b)$. All the other parameters are sampled with the priors listed in Table~\ref{tab:paramstraining}. In one particular case, that is to reproduce the results of \citet{Reichardt2020}, all cosmological parameters are fixed to the fiducial values shown in the same table.

\section{Results}\label{sec:results}

We use our modified version of \texttt{CosmoMC}, including the \Planck21-tSZ and the SPT high-$\ell$ likelihoods, to sample cosmological and scaling relation parameters and study the impact of updating the tSZ computation from an amplitude and a template to a physically motivated model.

\subsection{Impact of the RF-reconstructed tSZ power spectrum}

As a first step in our analysis, we
validate our machine-learning technique at the cosmological inference level. We compare the results of running \texttt{CosmoMC} with a full computation of the tSZ $C_\ell$'s, and with the RF-inferred ones. We compare in Fig.~\ref{fig:verif} the contours obtained in a MCMC run using our \Planck-tSZ likelihood  and varying two cosmological parameters ($\Omega_m$ and $\sigma_8$) and two scaling relation parameters, $\alpha$ (with a prior) and $(1-b)$. Results from the two approaches are fully in agreement.
Therefore, in the rest of this work, we rely on the RF-inferred $C_\ell$'s instead of the full derivation when using our analytical tSZ power spectra. They are referred throughout as `RF tSZ' power spectra. 

To further speed up the calculation, we also approximate the noise part coming from the full connected trispectrum by a constant term, as discussed in Sec.~\ref{sec:Planck_tSZ}. 
 
We report the results for the RF tSZ power spectrum with the constant trispectrum term in Fig.~\ref{fig:verif}, and show that
the approximation is consistent with the full evaluation. Since the trispectrum term becomes negligible at multipoles higher than 2~000, the range covered by the SPT data, this approximation will hold even better in the SPT analysis

\subsection{Impact on cosmological parameters}

\begin{figure*}
   \centering
   \includegraphics[width=0.5\textwidth]{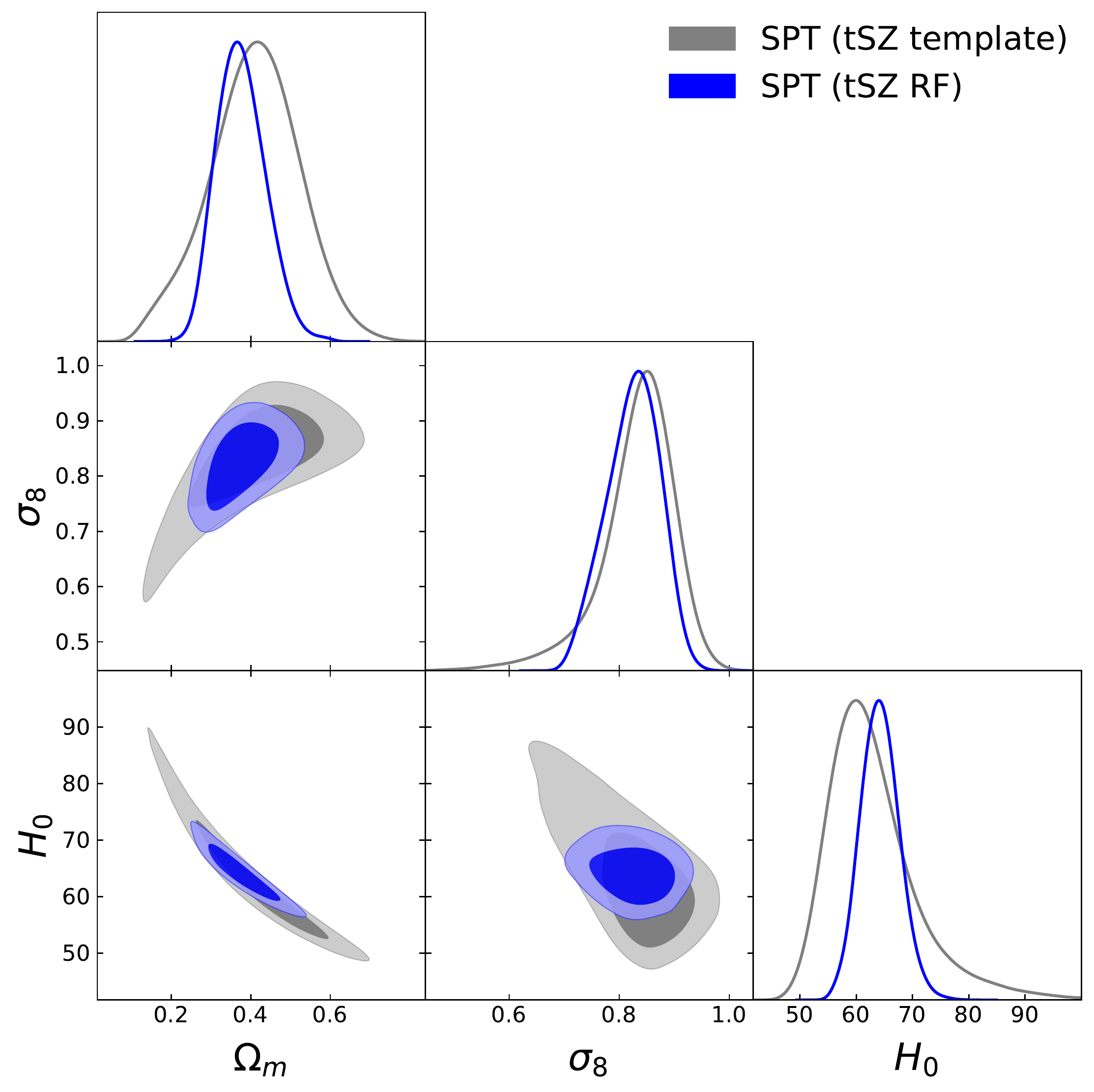}
       \caption{Constraints on cosmological parameters obtained with SPT high-$\ell$ data using the tSZ template as done in \citet[][in grey]{Reichardt2020} and using the tSZ power spectra derived from the halo model (in blue). } 
       
         \label{fig:cosmo_tempvshalo}
   \end{figure*}

We now study the consequences of substituting the tSZ template with cosmology-dependent spectra. We fitted the SPT data, letting the five cosmological parameters free with only priors on $n_s$ and $\Omega_b h^2$. We assume priors on the tSZ parameters $\alpha$ and $\log Y^*$, but leave the mass bias free to vary. Indeed, as shown in previous tSZ studies \citep{Planck13,Horowitz2017,Salvati2018,Bolliet18}, the cosmological parameters and the mass bias are degenerate, so that fixing the mass bias will be equivalent to setting a prior on the cosmological parameters. The priors used are listed in Table~\ref{tab:paramstraining}.
In practice, we have one more degree of freedom compared to \citet{Reichardt2020}: $\alpha$ and $(1-b)$ are added to describe the tSZ signal, whilst the amplitude parameter $A_{\text{tSZ}}$ at $\ell=3000$ is removed. In total, including foregrounds and SPT instrumental parameters, we have 22 free parameters. Nevertheless, as the shape of the tSZ power spectrum is now fitted to the SPT data, the constraints on cosmological parameters improve: Fig.~\ref{fig:cosmo_tempvshalo} compares the posterior distributions of $H_0$, $\Omega_m$ and $\sigma_8$ obtained with the template and with the RF. We can see that using SPT data alone with the tSZ template (in grey) allows for a large range of parameter values, including a narrow degeneracy between $H_0$ and $\Omega_m$. These constraints are driven by the CMB spectrum dominating other signals on the range $2000 < \ell < 4000$ but being negligible at smaller scales. As expected, substituting the template by our RF tSZ power spectrum (blue contours) allows one to exploit the full cosmological information, and the constraints are largely tightened with a factor of about two in improvement (see Table~\ref{tab:params_results}).

\begin{figure}
  \centering
  \includegraphics[width=\columnwidth]{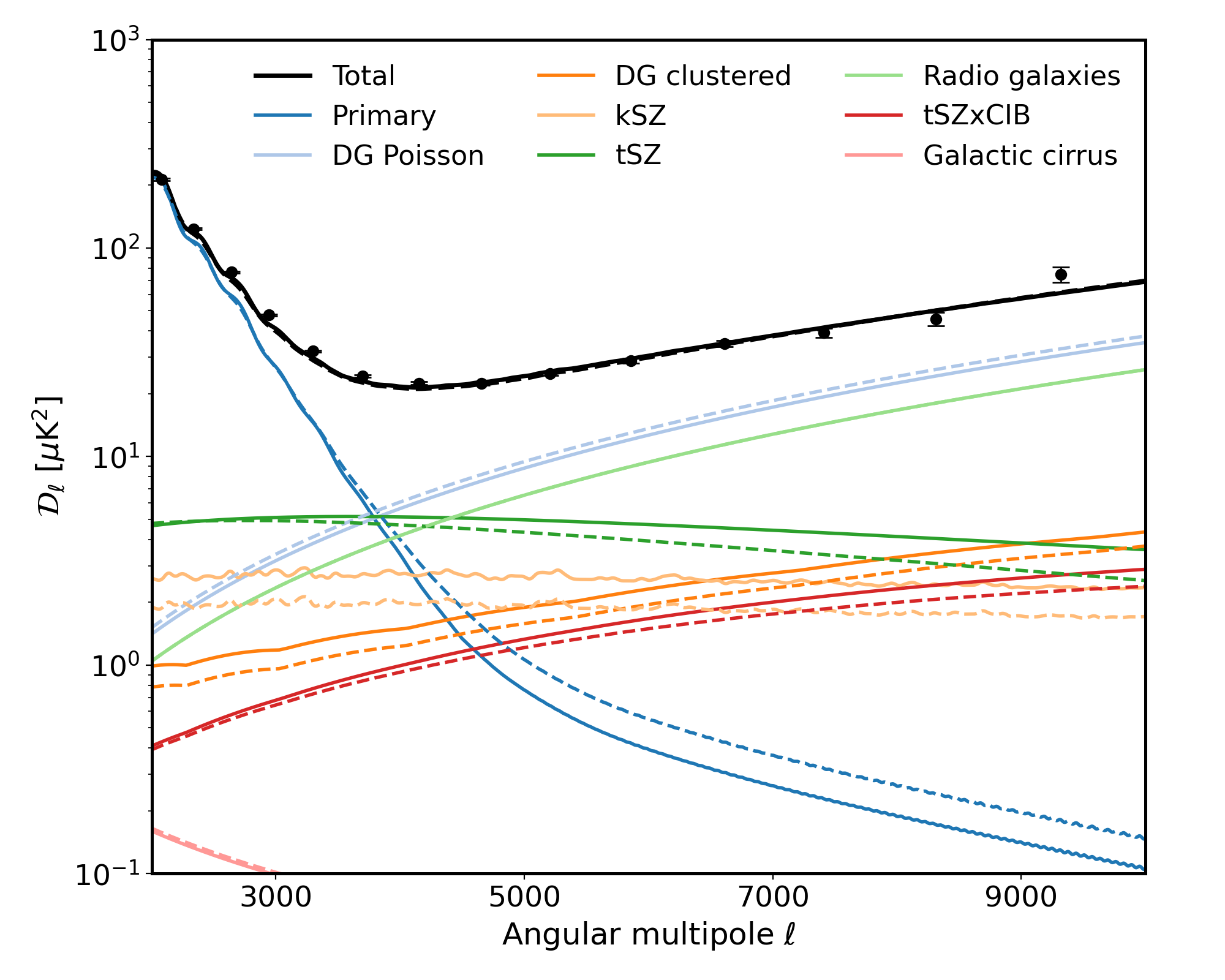}
       \caption{Best-fit spectra of SPT data for the 95x150~GHz bandpower (black data points) for CMB, tSZ and other foregrounds when using RF (solid lines) or a template (dashed lines) to derive the tSZ contribution.}
         \label{fig:cls_spt_fit}
  \end{figure}

Figure \ref{fig:cls_spt_fit} shows the best-fit spectra for the data at 95x150~GHz in the two scenarios (dashed lines for template, and solid line for RF), plotted as $D_{\ell} = \ell (\ell+1) C_{\ell}/ 2 \pi$. One can see the difference in amplitude and shape of the two tSZ spectra. 
Results for all SPT frequency bands and cross-spectra are shown in Appendix~\ref{app:additional_figures}.
Relative amplitudes of tSZ and kSZ are discussed in Sec.~\ref{sec:kSZ}.

\subsection{Combining Planck tSZ power spectrum and SPT data}
\label{subsec:res_SPT+planck}

\begin{figure}
  \centering
  \includegraphics[width=\columnwidth]{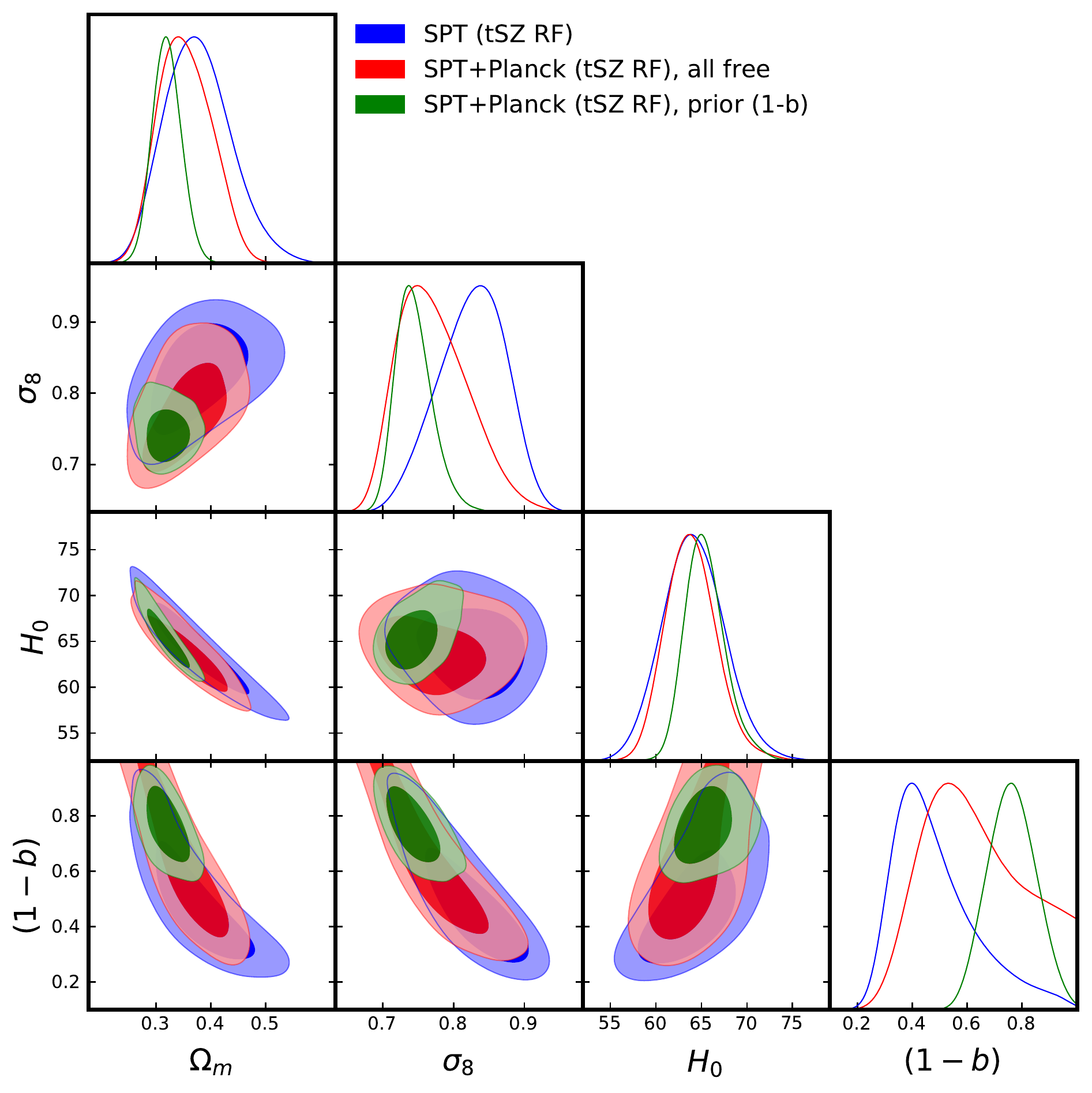}
           \caption{Constraints on cosmological and scaling relation
           parameters using different data sets and priors: in blue, SPT high-$\ell$ data alone, and in red, with \Planck21-tSZ power spectra added. Results when CCCP prior on the mass bias is added are shown in green. }
         \label{fig:cosmoparcons}
\end{figure}

In the following, we combine the \Planck21-tSZ power spectrum at large scales ($60<\ell< 1400$) with the SPT data at small scales ($\ell>2000$), described in Sec.~\ref{sec:likelihood}. We consider the case with RF power spectrum only, and compare with results for SPT data alone in Fig.~\ref{fig:cosmoparcons}. 
Adding the \Planck21-tSZ data (in red) only slightly tightens the constraints but moves the best-fit towards smaller values for $\Omega_m$ and $\sigma_8$  by roughly $1\sigma$.
These values are in better agreement with the \Planck~CMB constraints, within $1\sigma$ \citep{Planck2018_cosmo}. 

\begin{figure}
  \centering
  \includegraphics[width=\columnwidth]{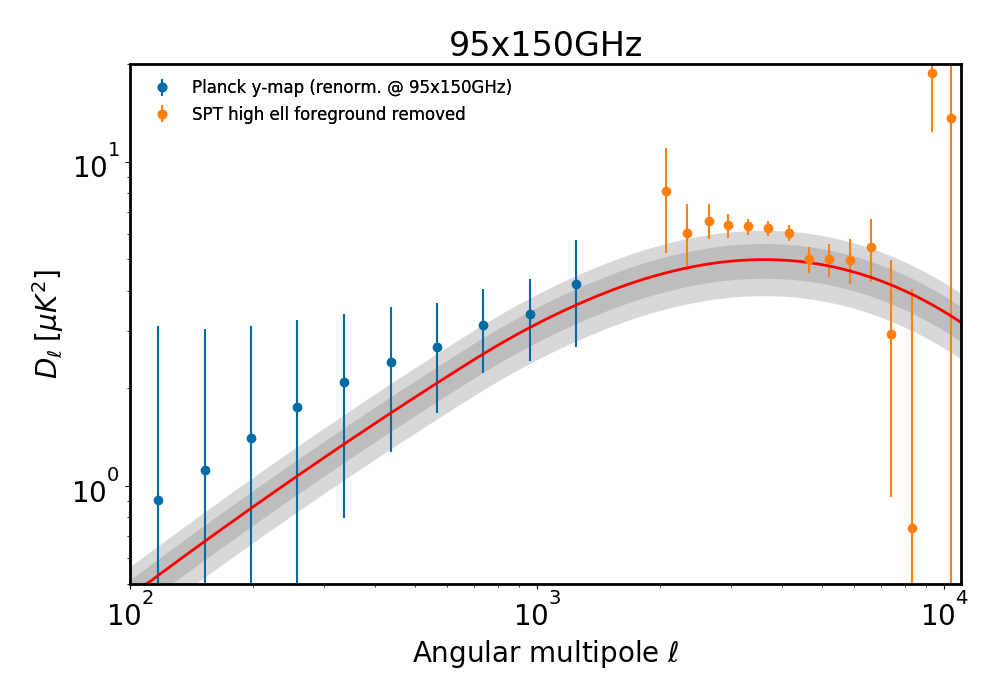}
      \caption{Best-fit SZ spectra of SPT (in orange) and \Planck21-tSZ (in blue) data,  foregrounds removed, at 95x150~GHz, when both of them are fitted simultaneously. The grey bands show the 1 and 2 $\sigma$ uncertainty on the best model. Similar plot for all cross spectra are shown in Appendix~\ref{app:additional_spectra}.}
         \label{fig:data_spt+plck_fit}
  \end{figure}

The most likely value for the mass bias $(1-b)$ similarly shifts 
towards a higher value when adding the \Planck21-tSZ data, now in better  agreement with the combination of \Planck\ CMB and \Planck\ SZ cluster counts \citep{Salvati2018}, but lower than most hydrostatic mass bias estimates (see \citet{douspis2019} and \citet{Gianfagna2021} for  compilations).
Adding a prior on the mass bias from the Canadian Cluster Comparison Project \citep[CCCP,][]{Hoekstra2015} breaks the degeneracy between cosmological and scaling relation parameters,  and eliminates high values of $\Omega_m$ and $\sigma_8$ (in green on Fig.~\ref{fig:cosmoparcons}). All results are reported in Table~\ref{tab:params_results}. 
Figure \ref{fig:data_spt+plck_fit} shows the tSZ power spectrum estimates of SPT at 95x150GHz once other foregrounds and primordial CMB are removed, as well as of \Planck21-tSZ estimates renormalised at the same frequency, along with our best-fit result. 

These 1-$\sigma$ best-fit changes shown in Fig.~\ref{fig:cosmoparcons} illustrate that the exact value of the best-fit depends on the assumptions on the foregrounds and calls for a more coherent analysis of large and small scale data, primordial and secondary anisotropies.

\subsection{Impact on kSZ detection }\label{sec:kSZ}

\begin{figure}[!t]
   \centering
   \includegraphics[width=0.45\textwidth]{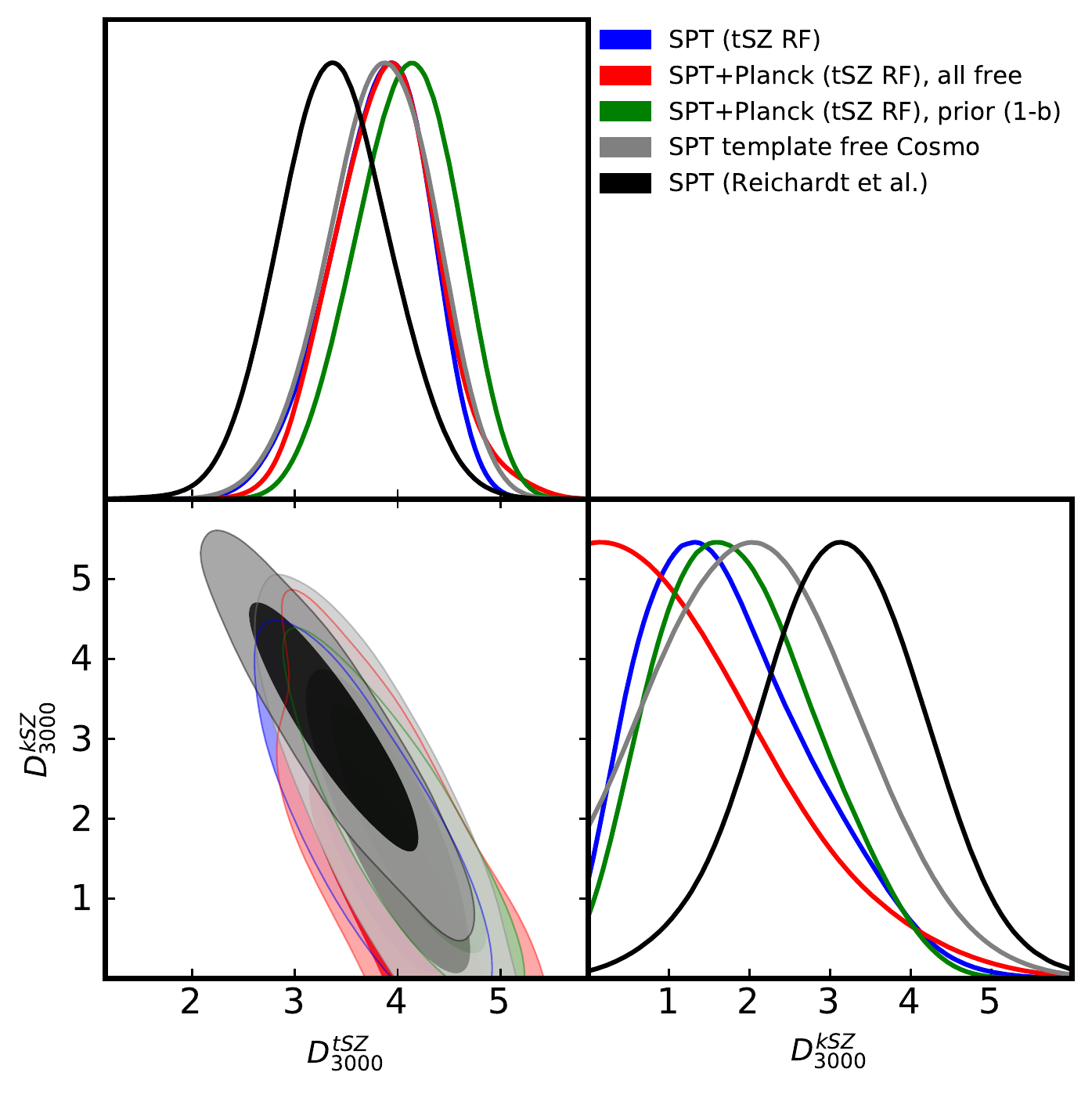}
       \caption{Constraints on the amplitude of the tSZ and kSZ power spectra at $\ell=3000$ and $143~\mathrm{GHz}$, using different data sets and methods to account for the tSZ signal. In grey and black are shown results obtained when using templates \citet{Reichardt2020}, with and without varying cosmology, respectively. Other colours refer to results using the halo model to derive the tSZ power and freeing cosmology. In blue, only SPT data is considered. In green and red, \Planck21-tSZ data is added, with Gaussian and uniform prior on the bias parameter, respectively. $D_{3000}^{\mathrm{kSZ}}$ and $D_{3000}^{\mathrm{tSZ}}$ are given in $\mu\mathrm{K}^2$. }
         \label{fig:tkszparcons}
   \end{figure}

\cite{Reichardt2020} have set constraints on the amplitude of the kSZ effect in SPT high-$\ell$ data, finding a 3-$\sigma$ detection when fixing the cosmology. The authors find that changing the templates used for the tSZ and kSZ power spectra makes the significance of their results range between 2.4 and 3-$\sigma$. We compare results on the kSZ power amplitude $D_{3000}^\mathrm{kSZ}$ in the different cases considered above, that is with fixed or free cosmology, with Gaussian or uniform priors on the cluster parameters. Figure~\ref{fig:tkszparcons} shows the one-dimensional likelihood on the kSZ amplitude for all cases, as well as the degeneracy between kSZ and tSZ amplitudes\footnote{estimated at $143~\mathrm{GHz}$} at $\ell=3000$. The default analysis by \citet{Reichardt2020}, using tSZ and kSZ templates, in black, gives the highest kSZ amplitude with the highest significance. Allowing, within the same set-up, cosmological parameters to vary\footnote{The priors on $n_s$ and $\Omega_b h^2$ presented in Table~\ref{tab:paramstraining} are used.} lowers the value of $D_{3000}^\mathrm{kSZ}$ by about $1\sigma$ (in grey). Substituting the template with the RF tSZ power spectrum leads to  different results than \citet{Reichardt2020}. Using the RF and freeing the cosmology allows the shape of the tSZ spectrum to change and thus better fit the details of the data. As a consequence, the kSZ amplitude and detection significance decrease drastically (in blue). The corresponding tSZ and kSZ spectra are shown in both scenarios in Fig.~\ref{fig:cls_spt_fit}, for the 95x150~GHz data. Finally, adding \Planck~tSZ data and prior on the mass bias moves the constraints by less than $1\sigma$ (in red and green, respectively).

These results show the sensitivity of the detection of kSZ effect to the assumptions made on other foregrounds or on the cosmological model used in the analysis. They motivate a similar treatment of other foregrounds such as the kSZ power spectrum and the CIB to ensure a robust detection of both SZ signals. The former is the subject of a companion paper \citep*{PAPER3}, focusing on kSZ and reionisation history.

\begin{table*}
    \centering
\begin{tabular}{c|c|c|c|c}
 & SPT (tSZ template) & SPT (tSZ RF) & SPT + Planck (tSZ RF), all free & SPT + Planck (tSZ RF), prior $(1-b)$ \\ 
 \hline
 \horsp
 $\Omega_m$ & $0.40 \pm 0.11$ & $ 0.38 \pm 0.06$ & $ 0.36 \pm 0.05$ & $ 0.32 \pm 0.02$ \\
 \hline
 \horsp
 $\sigma_8$ & $0.83 \pm 0.07$ & $ 0.825 \pm 0.048$ & $0.774 \pm 0.049 $& $0.744 \pm 0.025$\\
 \hline
 \horsp
 $H_0$ & $63.2 \pm 8.12$ & $64.2 \pm 3.4$ & $ 63.8 \pm 2.8$ & $65.4 \pm 2.2 $\\
 \hline
 \horsp
 $(1-b)$ & \textbf{-} & $ 0.49 \pm 0.16$ & $ 0.62 \pm 0.18$ & $0.76 \pm 0.09$\\
 \hline
\end{tabular}
    \caption{Constraints on cosmological and mass bias parameters for the different dataset combinations. We report the $68\%$ confidence levels. }
    \label{tab:params_results}
\end{table*}

\section{Conclusions}

Small-scale millimetre wavelength signals encompass a wealth of cosmological information. Several studies have used such information independently, focusing on the derived tSZ spectrum from $y$-maps \citep{Planck13,Planck2015_tsz_map, Salvati2018, Bolliet18}, the kSZ amplitude \citep{Zahn12, Planck2015_reion, Reichardt2020} or the measured CIB spectrum \citep{Planck14_cib, Maniyar21}. In contrast, in our framework, information is extracted from all cosmological signals at small scales simultaneously and coherently. 

In this first work, we focus on the tSZ power spectrum and how, in combination with CMB observations at small scales, it can help constraining cosmological parameters. Instead of assuming a fixed template shape and a free amplitude, we compute the tSZ angular power spectrum from the halo model from a set of cosmological and scaling relation parameters. However, this derivation is computationally intensive and we choose to train a Random Forest (RF) on 15~000 of tSZ angular power spectra and use the resulting predictions (RF prediction code for tSZ power spectra  available at \url{https://szdb.osups.universite-paris-saclay.fr/}), accurate at more than the 2\% level, in a full Markov Chain Monte Carlo analysis of the SPT high-$\ell$ data \citep{Reichardt2020}. Considering the width of current error bars on small-scale measurements of the CMB power spectrum and the modelling uncertainties of other CMB foregrounds, such an accuracy level is sufficient, but will need to be improved once data from the next generation of CMB observatories is available. We improve the constraints on  $\Omega_m$, $H_0$ and $\sigma_8$  by a factor of six compared to the case neglecting tSZ cosmological information, considering the figure of merit $\text{FOM}=1/(\sigma_{H_0}\sigma_{\Omega_m}\sigma_{\sigma_8})$. This represents the first step towards using the full potential of small-scale CMB data.  
 We then combine the SPT high-$\ell$ data with the \Planck21-tSZ power spectrum \citep{Tanimura2021} to further constrain cosmological and scaling relation parameters. This combination does improve constraints but mostly shifts the best-fit parameters towards smaller values of $\Omega_m$, and $\sigma_8$ (from 0.38 and 0.82 to 0.35 and 0.77, respectively), in better agreement with \Planck~CMB data. Adding a prior on the mass bias parameter from CCCP tightens the constraints on the three cosmological parameters of interest by another factor of six and brings the best-fit towards those of other tSZ analyses for a value of $S_8 \equiv \sigma_8 (\Omega_m/0.3)^{0.5}=0.76\pm 0.04$, while remaining compatible with \Planck~CMB results \citep{Planck18_cosmo}.
 Conversely, combining high-$\ell$ and tSZ data with large-scale CMB \citep{Planck18_cosmo} or future small-scale experiments \citep[][respectively SPT3G, Simons Observatory and CMB-S4]{SPT3G,Simons,S4} will constrain the cosmological model strongly and, in turn, the scaling parameters. Such constraints on a mean scaling relation, averaged over all masses and all redshifts, would need to be compared with most recent estimates using large samples of clusters in X-ray \citep{XLSS,Lovisari20,eRosita,eROSITA-mass} or in optical \citep{Sartoris16,Salvati20,Schrabback21,Roman}. We choose to limit ourselves to qualitative comparisons.  Indeed, we expect constraints to change once the cosmological dependence of other foregrounds, such as the kSZ and the CIB, as well as large-scale data, are included in the analysis.
 
 Finally, we investigate the amplitude of the deduced kSZ contribution and the amplitude of the tSZ effect on small scales ($\ell=3000$). When compared with the original study by \citet{Reichardt2020}, which fixed cosmological parameters and used templates for all small-scale foregrounds, this new analysis leads to a lower contribution of the kSZ to the overall observed power. Its amplitude is still detected but with decreased significance $D_{3000}^{kSZ} = 2.1\pm1.1~\mu \mathrm{K}^2$.
A more careful analysis, also including the cosmological dependency of the kSZ and CIB effects, needs to be performed to obtain a more robust estimate of the contribution of the kSZ effect to small-scale CMB data and its implications in terms of the Epoch of Reionisation. This is the focus of a forthcoming companion paper \citep*{PAPER2}. 

Whilst more complete and coherent, our analysis is based on a model giving the tSZ power spectrum from a set of parameters and is, therefore, limited by the physical assumptions made for modelling the astrophysical contributions to the small-scale power. These limitations should be overcome, and the accuracy of the RF prediction improved, once data from future surveys -- or their combination, with enough sensitivity to constrain both cosmological and astrophysical parameters simultaneously, is available.

\begin{acknowledgements}
The authors acknowledge the SPT collaboration and C. Reichardt for making the SPT data and likelihood available online\footnote{Likelihoods, figures and data points from \citet{Reichardt2020} are available at \url{https://pole.uchicago.edu/public/data/reichardt20/}.}.

AG's work was supported by the McGill Astrophysics Fellowship funded by the Trottier Chair in Astrophysics, as well as the Canadian Institute for Advanced Research (CIFAR) Azrieli Global Scholars program and the Canada 150 Programme. \\
NA is partly funded by the ByoPiC project from the European Research Council (ERC) under the European Union's Horizon 2020 research and innovation programme grant agreement ERC-2015-AdG 695561.
The authors acknowledge fruitful discussions with the members of the ByoPiC project (https://byopic.eu/team).\\
This research made use of the computation facility of IDOC (http://idoc.ias.u-psud.fr), partly provided by DIM ACAV (http://www.dimacav-plus.fr).
This research made use of \texttt{matplotlib}, a Python library for publication quality graphics \citep{hunter_2007}, \texttt{Astropy}, a community-developed core Python package for Astronomy \citep{2013A&A...558A..33A}, of \texttt{scipy}, a Python-based ecosystem of open-source software for mathematics, science, and engineering \citep{scipy} -- including \texttt{numpy} \citep{van2011numpy}, and of \texttt{scikit-learn} \citep{scikit-learn}.
\end{acknowledgements}

\bibliographystyle{aa} 
\bibliography{biblio} 

\appendix

\section{Parameters}\label{app:params}

\begin{table*}[!h]
    \centering
    \small
\begin{tabular}{lllll}
Definition & Symbol & Prior & Fiducial &  Sampling range\\ \hline \hline
Baryon density & $\Omega_b h^2$ & $0.0224\pm0.00015$  & 0.224 & [0.01:0.03]       \\
Dark Matter density & $\Omega_c h^2$ & -   & 0.1193 & [0.1:0.3] \\
Amplitude  of fluctuations & $\log A$ &  - & 3.047 &  [2.0:4.0]      \\
Spectral index   &   $n_s$     &  $0.9649\pm0.0044$  &0.9665& [0.92:1.02]     \\
CMB sound horizon to distance ratio & $\theta$ & - & 1.04101& [0.99:1.1]  \\ 
Reionisation optical depth & $\tau$ & 0.056 & 0.056 & Always fixed \\ \hline
Hydrostatic mass bias & $(1-b)$ & $0.78\pm 0.09$ & 0.78 &  [0.2,1.0]\\ 
SR mass slope    & $\alpha$  &  $1.79\pm0.08$ & 1.79 & [ 1.5:2.1]\\
SR amplitude    &  $\log Y^* $      & $-0.186\pm0.021$  & -0.186 & [-0.29:-0.08]\\
SR redshift slope & $\beta$ & 2/3 & 2/3& Always fixed \\ \hline
\end{tabular}
    \caption{Definition, symbol, prior and ranges assumed for considered parameters. `SR' stands for scaling relation.}
    \label{tab:paramstraining}
\end{table*}

Table~\ref{tab:paramstraining} lists all parameters, cosmological and scaling relation, related to the tSZ spectrum, along with their definitions, values when fixed, and applied priors.

\section{Best-fit spectra at different frequencies}
\label{app:additional_figures}

We reproduce here Fig.~\ref{fig:cls_spt_fit} but with a different colourmap and for all combinations of frequencies allowed by the data, that is auto-spectra at 95, 150 and 220~GHz, and cross-spectra at $95\times150$, $95\times220$ and $150\times220$~GHz.

\begin{figure*}[h]
  \centering
\includegraphics[width=\textwidth]{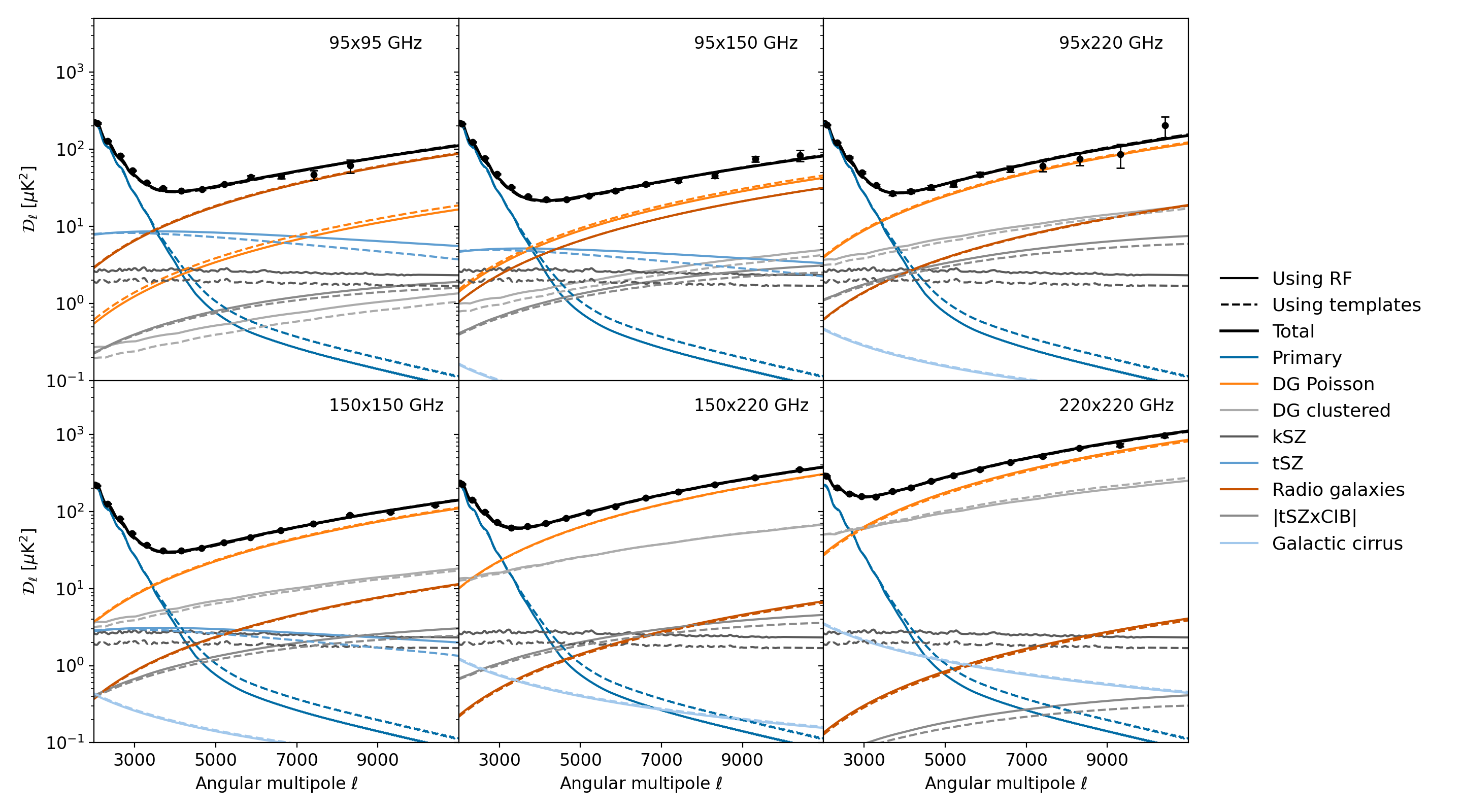}
      \caption{Best-fit spectra of SPT data for CMB, tSZ and other foregrounds, as well as the sum of all compared with the SPT data at different frequencies. We compare results when templates or RF predictions are used to obtain the tSZ power spectrum.}
         \label{fig:cls_all}
  \end{figure*}

\section{Best-fit SZ spectra estimates for all (cross) frequencies}
\label{app:additional_spectra}

We reproduce here Fig.~\ref{fig:data_spt+plck_fit} for all combinations of frequencies allowed by the SPT data, that is auto-spectra at 95, 150 and 220~GHz, and cross-spectra at $95\times150$, $95\times220$ and $150\times220$~GHz.

\begin{figure*}[h]
  \centering
\includegraphics[width=\columnwidth]{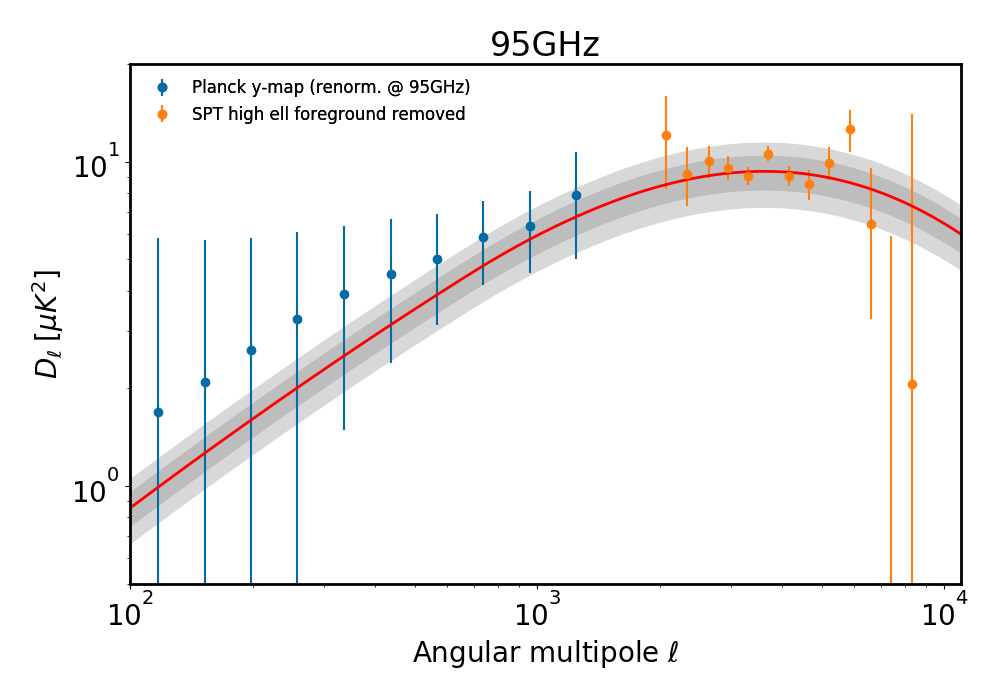}\includegraphics[width=\columnwidth]{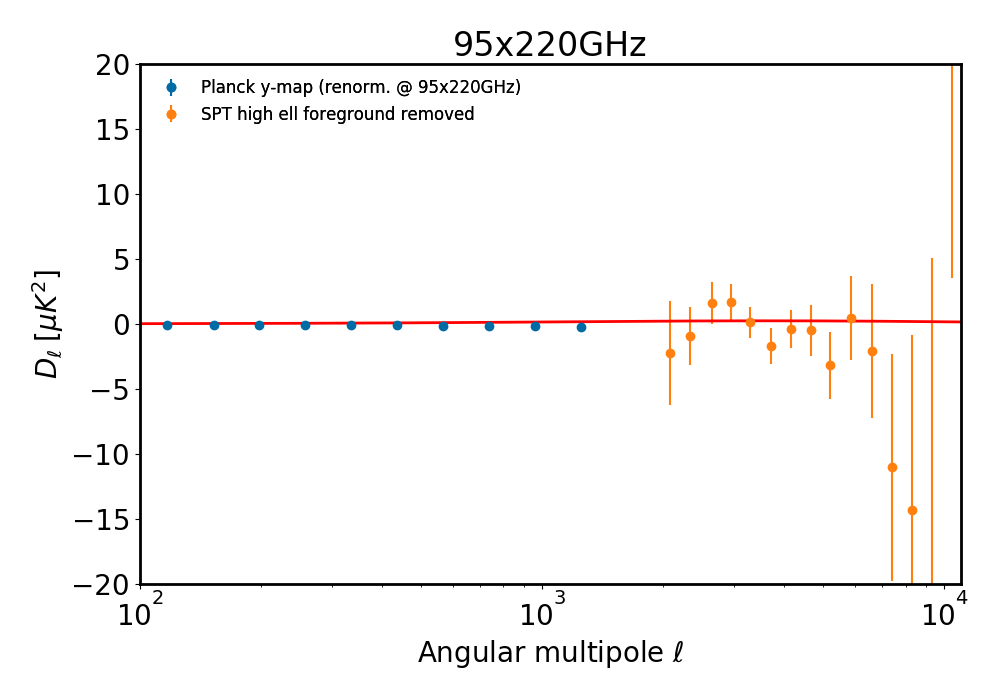}
\includegraphics[width=\columnwidth]{Figs/Fig2021_dataCls_95x150.png}\includegraphics[width=\columnwidth]{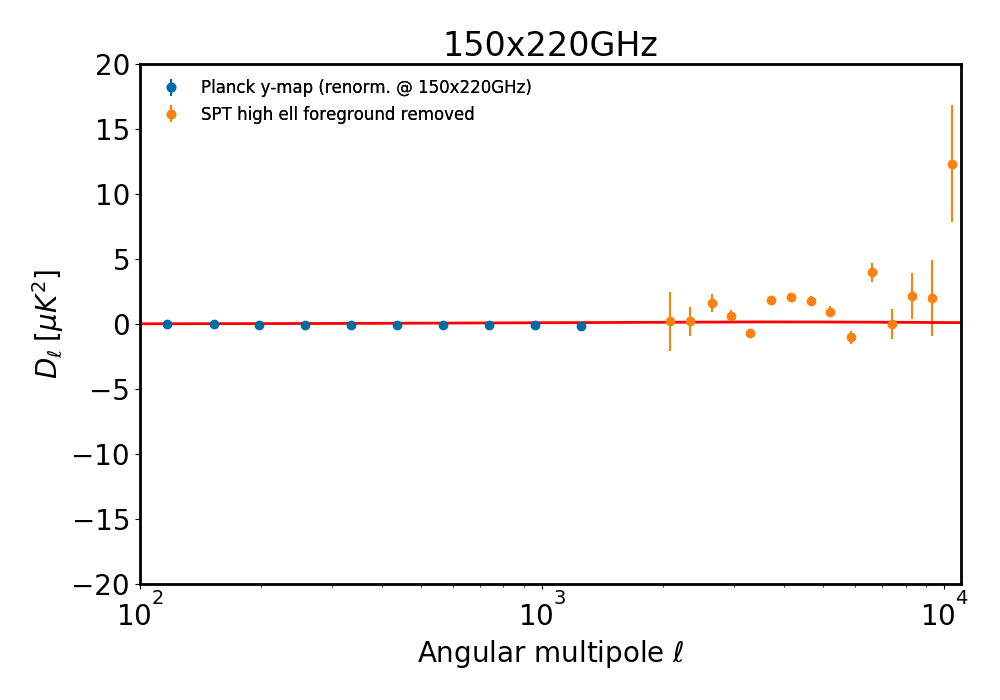}
\includegraphics[width=\columnwidth]{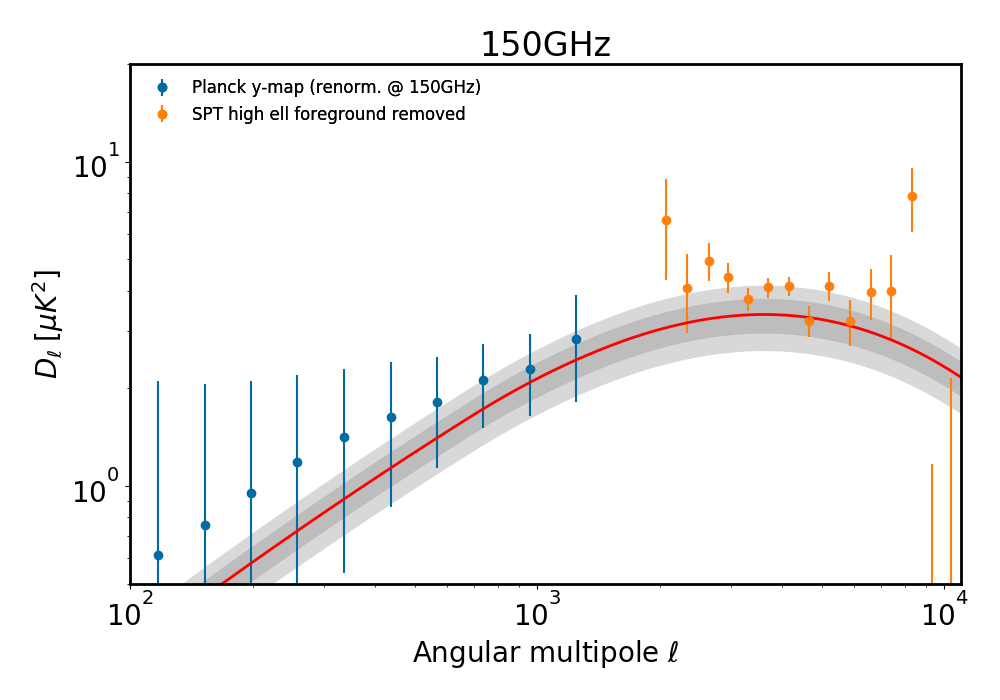}\includegraphics[width=\columnwidth]{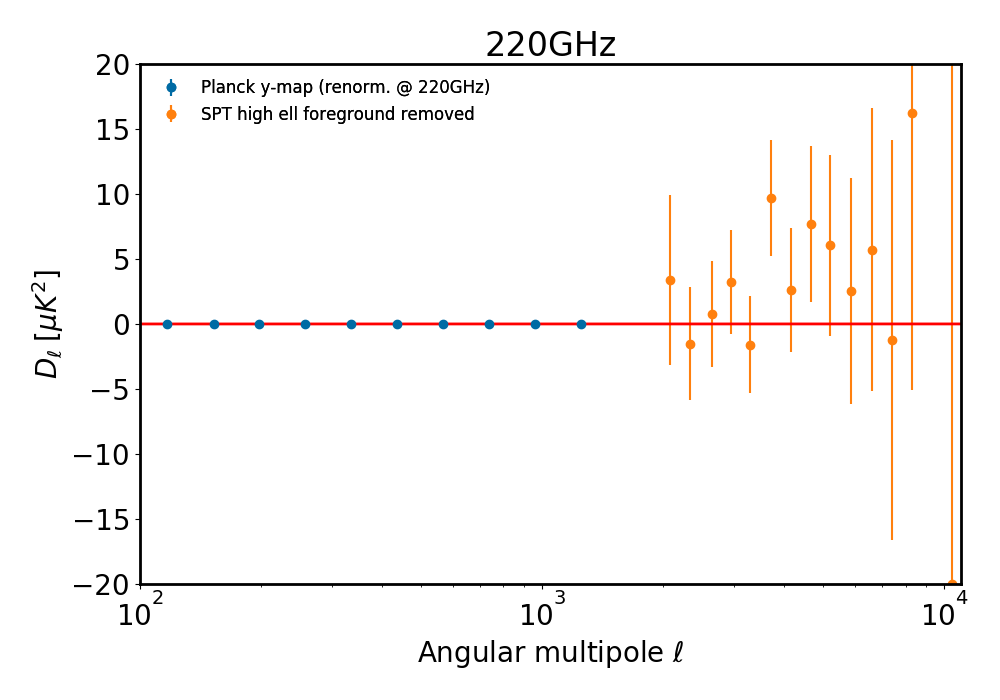}
       \caption{Same as Fig.~\ref{fig:data_spt+plck_fit} but at different frequencies}
         \label{fig:clsdata}
  \end{figure*}

\end{document}